# Calculation principles for a superconducting inductive FCL and a current-limiting transformer

Vladimir Sokolovsky, Victor Meerovich,

*Department of Physics, Ben-Gurion University of the Negev, P. O. Box 653, Beer-Sheva,*

Istvan Vajda,

*Department of Electric Power Engineering, Budapest University of Technology and Economics, Budapest, H-1111 Hungary.*

*Abstract*

We present general principles of the parameters calculation and application of two power devices: an inductive fault current limiter and a current-limiting transformer. Both the devices are based on the superconducting – normal state transition in a superconducting switching element magnetically coupled with a primary winding. The relationships between the design parameters of the devices and the limited current, losses and temperature in the superconducting element were investigated. It is shown that, in spite of similarity, the devices have distinctive features manifested in different depths of the fault current limitation. The case of performance of the superconducting winding in the form of HTS BSCCO hollow cylinders is analyzed. Basic parameters of the devices are evaluated for their application in 45 MVA power circuit. Problems of the transition into the normal state and recovery of the superconducting state in the switching element are discussed.

*Index terms* - superconductors, fault current limiter, current-limiting transformer

## I. Introduction

The reduction of the damage caused by high short-circuit currents is one of vital problems of power engineering. To reduce fault currents, different fault current limiters (FCLs) were proposed, among them devices which are based on a fast transition from the superconducting (SC) state to the normal state (S-N transition). One design - a limiter with the magnetic coupling between an SC element and a protected circuit, so called

"inductive FCL" - was investigated by many authors for various parameters and performances of an SC element [1-4]. Typically this device is a two-winding transformer with an SC short-circuited secondary winding and a primary winding inserted in series into the circuit to be protected. The limitation effect is achieved as a result of the impedance increase of an FCL when any part or the whole SC winding passes into the normal state. The secondary winding in an inductive FCL employing high-temperature superconductors (HTS) is performed frequently in the form of a set of HTS rings or cylinders. As an alternative design, multiple-turn SC coils short-circuited by SC switching elements were considered [5,6]. In the last case only the switching element undertakes the S-N transition at a fault while the secondary coil remains in the SC state. Several high power prototypes of an inductive FCL have been built and successfully tested showing feasibility of the concept for the application in power electric systems [4, 7-10].

The inductive design was also used as a basis for the development of more complicated power devices combining the current-limitation with other functions [11-15]. One of them is a current-limiting transformer (CLT) (it is known also as a transformer-circuit-breaker) which is capable to limit fault currents in the circuit connected to its secondary winding to a very low level. This device was first proposed by the authors of [11]. Experimental investigations performed with models of the CLT have shown the feasibility of this design [12-15].

In the present paper we consider two devices providing current limitation under faulty conditions: an inductive FCL and a CLT. The goal of the paper is to give the general principles of the parameters calculation for these devices and compare them in application examples.

**II. Inductive FCL: requirements to parameters and principles of calculation**

The device impedances under normal and fault regimes, activation current, times of the activation and initial state recovery are the basic "external" parameters of an FCL. The acceptable values of these parameters lie in the limits prescribed by characteristics of the normal and fault regimes of the power network [3].

An ideal limiter has zero impedance under the normal regimes of the circuit to be protected and changes its impedance quickly (before the first peak of a fault current) from zero up to a reasonable large value required for the limitation of a fault current. In reality, it is enough that the voltage drop across an FCL under the normal conditions of a circuit is less than several percents (usually 5%) of the rated circuit voltage. In this case the FCL does not disturb the static stability of the power system operation and does not influence handling properties of the lines. An inductive FCL under the normal regime is a transformer with a short-circuited secondary SC winding. The impedance of the device in this regime is determined by the leakage inductances of the windings. When the windings are located on the same leg of the magnetic core, the inductive reactance is determined by the clearance $\delta$ between two coaxial windings, the thickness $b_1$ of the primary with the turn number $w_1$ (neglecting the thickness of the SC winding), the diameter $D$ and height $h_1$ of the winding:

$$x_s = \omega\mu_0 w_1^2 \pi D k_\rho (b_1/3 + \delta)/h_1 \qquad (1)$$

where $\mu_0$ is the vacuum permeability, $k_\rho$ is the Rogovski coefficient ($k_\rho \approx 0.9$) [3, 5. 16].

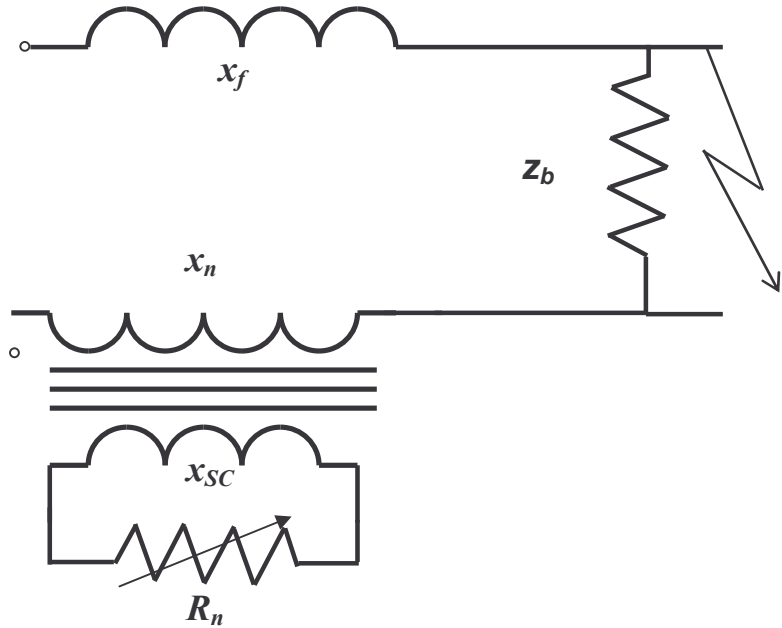

Fig. 1. One-phase equivalent circuit of a network with an inductive FCL: $x_f$ is the short-circuit impedance; $z$ is the load; $x_n$ and $x_{sc}$ are the inductive reactances of the primary and secondary (SC) windings; $R_n$ is the non-linear resistance appeared in the SC winding.

The clearance between the windings can be minimized by keeping them as close together as possible. However, the SC winding has to be placed into a cryostat with coolant, and the minimal clearance $\delta$ is determined by the design features of cryostat and the required electrical isolation. The thickness of the primary winding $b_1$ increases with the nominal voltage due to the increase of the electric isolation and with the nominal current due to the increase of the wire diameter. One of the ways to decrease the winding thickness and clearance and, hence, $x_s$ is to perform the primary winding from an SC wire and to place it in the cryostat together with the secondary SC winding.

The impedance of an FCL in the limitation regime is determined by both the inductive reactance and the resistance appeared in the SC winding (Fig. 1). In this section we will consider the more general case, when the SC winding contains a number of turns and is closed by an SC switching element, which undertakes the S-N transition. It is requested usually [3, 5] that the resistance of a switching element in the normal state $R'_n$ referred to the primary should be much larger than the inductive reactance of the primary winding $x_n$. Then the FCL impedance $z_n \approx x_n$ is close to the impedance of a transformer with an open-circuit secondary winding. However, this condition was not fulfilled in many successfully tested FCL prototypes which had a pronounced resistive component of the impedance [7, 8, 17, 18]. In this more general case, the steady-state limited current is determined by[1]

$$I_L = I_M \sqrt{1 + \left(x_n / R'_n\right)^2}. \tag{2}$$

Here $I_M$ is the magnetizing current (the current in the branch $x_n$ in Fig. 1) given by

$$I_M = \frac{U_{nom} R'_n}{\sqrt{x_n^2 x_f^2 + (x_n + x_f)^2 R'^2_n}} \tag{3}$$

where $U_{nom}$ is the nominal (phase) voltage of the power line, $x_f$ is the reactance of the circuit under a fault (short-circuit impedance).

In the following analysis, we neglect the transient process caused by the device activation and assume that the switching element resistance is changed stepwise from zero to $R'_n$.

---

[1] Here and below we use a simplified one-phase equivalent circuit of the network.

The loss power dissipated in the resistance $R'_n$ is determined as

$$P = \frac{x_n^2 I_M^2}{R'_n}. \qquad (4)$$

The dependencies of the limited current and loss power on the resistance $R'_n$ (in the base units) are shown in Fig. 2. One can see that already for $R'_n/z_b > 3$ the limited current is determined by $x_n$. It is possible to obtain the current limitation also with smaller values of $R'_n$ but herewith the losses considerably increase (Fig. 2b). For very small values of the resistance, the losses decrease but the limited current grows and practically does not depend on $x_n$.

The losses are a very important parameter of the FCL as they determine not only economic effectiveness of the device but also the feasibility of its operation. Due to the losses after the S-N transition the switching element has to be heated above the critical temperature and to keep the normal state during a fault event. Excess losses lead to overheating of the switching element and make difficulties the recovery of the initial SC state after fault clearing.

Due to a small time of a fault regime $t_{lim}$ (usually about 0.1 s), the maximum temperature of the heating can be evaluated as in adiabatic conditions:

$$\Delta T = P\, t_{lim}/C_V V \qquad (5)$$

where $C_V$ is the specific volume heat capacity, $V$ is the volume of the switching element. The cross-section area $s$ of the switching element is determined by the ratio of its critical current $I_c$ to the critical current density $J_c$. The required critical current of the switching element is calculated as

$$I_c = I_{act} \frac{w_1}{w_s} \qquad (6)$$

where $w_s$ is the turn number in the SC winding; $I_{act}$ is the activation current, i.e. is the instantaneous circuit current at which the SC switching element passes into the normal state. As a fault regime arises if the circuit current is higher than doubled maximum of the rated current in the normal regime [3, 19], it should be $I_{act} \geq 2\sqrt{2} I_{nom}$.

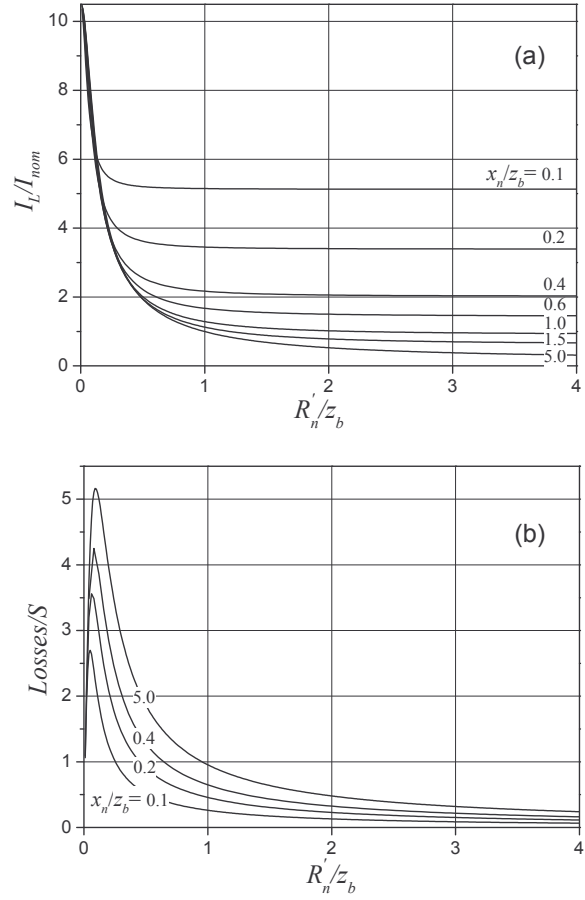

Fig. 2. Limited current (a) and losses (b) vs. the normal state resistance of a switching element (referred to the primary). $x_f/z_b = 0.095$. $I_{nom}$ is the nominal current in the circuit; the base units are $S = U_{nom}I_{nom}$; $z_b = U_{nom}/I_{nom}$.

Therefore, the required volume of the switching element is determined by the expression:

$$V = ls = \frac{R'_n I^2_{act}}{\rho J_c^2} \qquad (7)$$

where $\rho$ is the normal state resistivity of the superconductor.

Substituting (4) and (7) in (5) and choosing $I_{act} = 2\sqrt{2} I_{nom}$, we obtain

$$\Delta T = \frac{\rho J_c^2 t_{\lim}}{8 C_V} \left(\frac{I_m}{I_{nom}}\right)^2 \left(\frac{x_n}{R'_n}\right)^2. \qquad (8)$$

Thus, the temperature increase depends strongly on $R_n'$ (Fig. 3). For every value $x_n$, one can find a maximal value of $R_n'$ providing the heating of the switching element above the critical temperature. According to (8), this maximal value increases with $\rho J_c^2$, so that employing superconductors with higher $J_c$ one can increase $R_n'$ and, hence, decrease the losses.

The minimal allowable value of $R_n'$ is determined from the conditions of accessible heating of the switching element and recovery of the SC state after fault clearing during ~ 1 sec. The recovery time is determined by both the temperature increase $\Delta T$ and the cooling conditions which depend strongly on the design features.

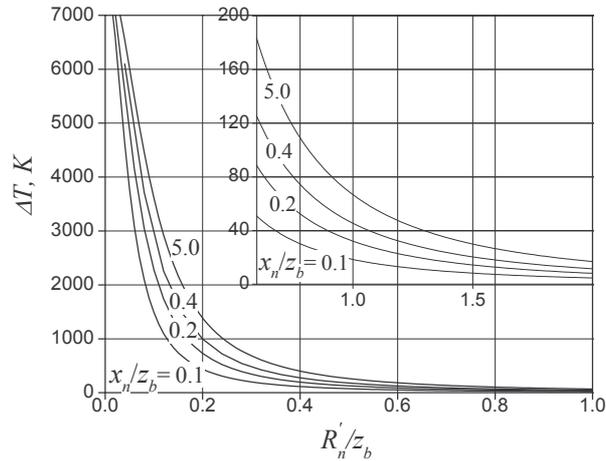

Fig. 3. Temperature increase of the SC element during the limitation time as a function of the normal state resistance. $x_f/z_b = 0.095$. The superconductor parameters are $\rho = 10^{-5}$ $\Omega\cdot$m, $J_c = 2\times10^7$ A/m$^2$, $C_V = 0.7\times10^6$ J/(m$^3$K).

Emphasize that the choice of the inductive reactance $x_n$ and resistance $R_n'$ is dictated by the required value of the limited current $I_L$ which is the only parameter interesting for power utilities. The range of possible values of $I_L$ is discussed in [3]. As noted, a usual requirement to an FCL installed in a power network is that the minimum of the limited fault current is greater than doubled nominal current $I_{nom}$. If a fault current is lower than $2I_{nom}$, the system automatic equipment cannot "see" a fault condition and does not command to the circuit breaker to open the circuit. However, this fact cannot be determining for the FCL parameters. It means only that, in the case of the deep limitation, the automatics should be rebuilt so that it operates when the FCL is activated.

The maximum accessible ratio $x_n/x_s$ determines the applicability of different designs of an inductive FCL. There are three basic designs of an electromagnetic system: 1) open magnetic core (a leg without yokes); 2) closed core with distributed gaps; 3) air reactor (without core). Estimates show that the maximum of $x_n/x_s$ for an air reactor is 6-8, for an open core design is 20 [3, 5]. Since $x_s$ is limited by 5% voltage drop in the normal regime, we cannot use an open core design with $x_n > 1$. To achieve a higher ratio, one should apply a closed core design. As an air reactor possesses a low ratio of the impedance change, we restrict ourselves to the study of the designs with a magnetic core. The expressions for calculation of the parameters of these designs are given in [3, 5, 16, 20-22]. The expression obtained in these papers refer to case when $R'_n \gg x_n$ and, hence, $I_M = I_L$. In the general case, $I_M \neq I_L$ and the core diameter and turn number in the primary winding are determined by:

$$D_{core} = K_1(x_n I_M^2)^{1/3}, \quad (9)$$

$$w_1 = K_2(x_n/I_M)^{1/3} \quad (10)$$

where $K_1$ and $K_2$ are the coefficients determined by the design features and admissible maximum magnetic flux density $B_{max}$ in the core ($B_{max}$ should be less than the saturation flux density, about 2 T).

For a closed core design, the coefficients $K_1$ and $K_2$ depend on the total air gap $\eta$:

$K_1 \sim \eta^{-1/3}$, $K_2 \sim \eta^{2/3}$.

To calculate the parameters $D_{core}$ and $w_1$, it is necessary to set the resistance of the switching element and limited current. For the closed core design, there is an additional variable parameter $\eta$. To determine $\eta$, one can use the restriction on the value $x_s$ given by (2): $x_s < 5\% z_b$.

As an example, we will give the dependencies of $D_{core}$, $w_1$ and $x_n$ on the limited current for the design with an open magnetic core (Fig. 4). It is of interest that, at a small ratio of $R'_n / z_b$, the curves have a double-valued character so that the same limited current can be achieved by two different sets of the FCL parameters.

The dependence of the core diameter on the limited current has a clearly defined maximum which does not depend almost on $R'_n$. The core diameter can be reduced by choosing a lower limited current (Fig. 4a) but herewith the turn number in the primary winding increases (Fig. 4b) resulting in the increase of the reactances $x_n$ and $x_s$. When the last achieves 5% limit, a further reduction of $I_L$ is inadmissible. The minimal limited current with an open core design is about $I_{nom}$.

Note that, according to (6), the increase in the turn number $w_1$ leads to the increase of the critical current of the switching element. To provide the same critical current, one has to increase $w_s$ proportional to $w_1$.

The relationships presented above form the groundwork for calculation of the parameters of an inductive FCL and allow to analyze the application area of the device.

**III. Current-limiting transformer**

*A. Design and operation principle*

The device proposed in [11], so called "current-limiting transformer" (CLT), constitutes a single-phase transformer with two additional windings located on the same leg (Fig. 5): one of them is an SC short-circuited winding, the second is counter-connected in series with the main secondary winding so that the voltages induced in the windings are of opposite signs. The numbers of turns in the windings $w_2$ and $w_3$ can be chosen so that with opened SC winding their induced electromotive forces are compensated. Fig. 5 presents one possible arrangement of the windings: a few others are described in [11, 13, 14].

As and for an FCL, two regimes of the operation have to be considered. Magnetic flux $\Phi_1$ in the central leg is determined by the rated primary voltage $U_{nom}$ : $\Phi_1 = U_{nom}/\omega w_1$ ($\omega$ is the cyclic frequency of current) and remains the same in all the regimes. Under the normal conditions of the network, the SC short-circuited winding compensates the magnetic flux in the right leg (Fig. 5). Therefore the voltage on winding $w_3$ is zero. The same flux $\Phi_3 = \Phi_1$ exists in the left leg. The device operates as a usual one-phase transformer.

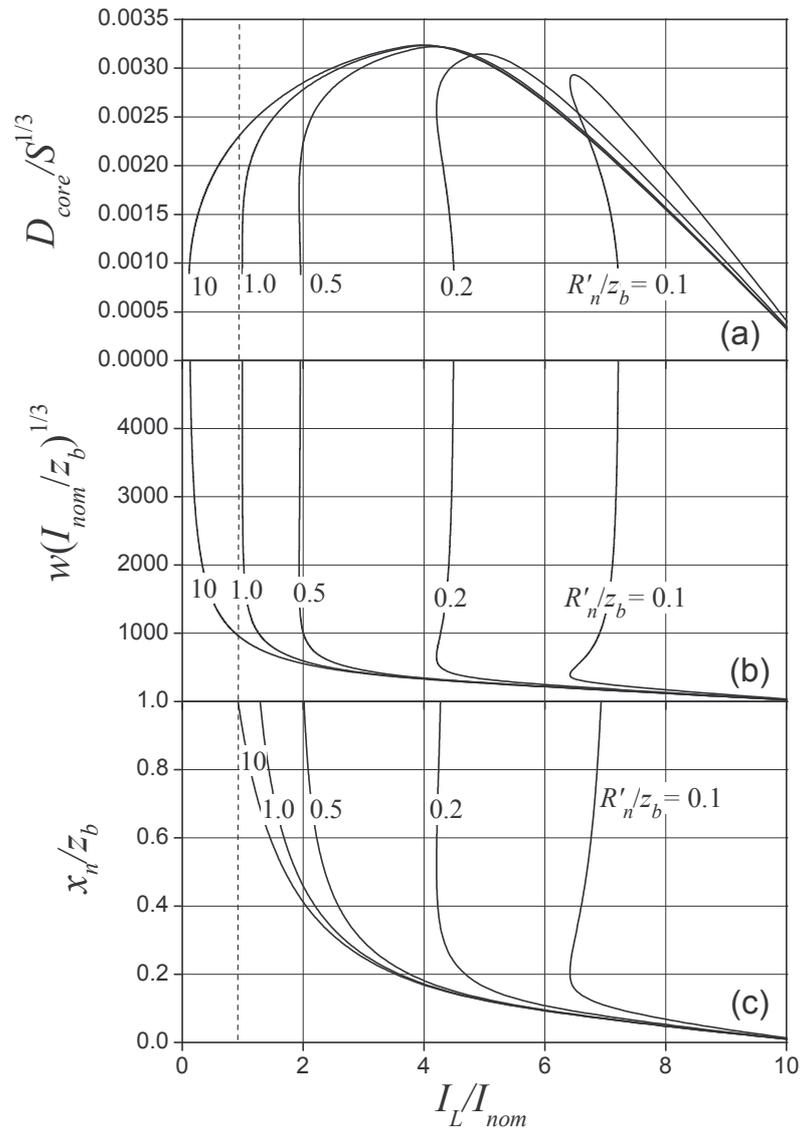

Fig. 4. Magnetic core diameter (a), turn number of the primary winding (b) and inductive reactance (c) as functions of the limited current for different $R'_n / x_n$. Design with an open magnetic core, $K_1 = 0.00219$, $K_2 = 937$, (SI units) – for voltage class of 10-35 kV. $x_f/z_b = 0.095$. The dashed line marks the minimum accessible limited current for an open core design.

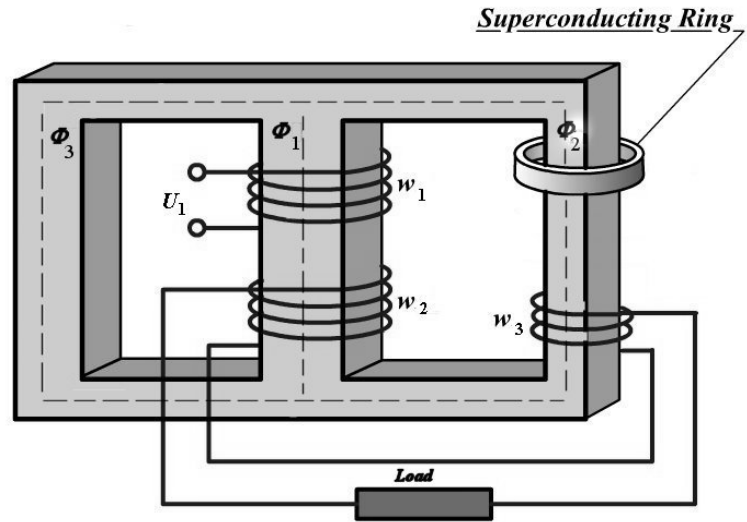

Fig. 5. Schema of a current-limiting transformer: primary winding is $w_1$, the secondary is formed by windings $w_2$ and $w_3$ counter-connected in series.

Under the faulty conditions, the increase of the current in the secondary circuit leads to the increase of the current in the SC winding and it passes into the normal state. The magnetic flux penetrates now into the right leg. As two sections of the secondary windings are counter-connected, the secondary voltage decreases drastically and the fault current is deeply limited. While the FCL action is based on inserting additional impedance in a circuit, the limiting property of a CLT is provided by the compensation of the voltage in the secondary circuit.

The transformer can be inversed, i. e. the two-sectional winding $w_2$-$w_3$ can be a primary winding and winding $w_1$ as a secondary. Also the device can be performed also as an autotransformer with the zero point in the connection between $w_2$ and $w_3$. Windings $w_1$, $w_2$ and $w_3$ can be performed from normal metal or SC wire. So, if the CLT is used as an input transformer for an SC cable, the primary winding is normal metal and the secondary windings are SC.

*B. Operation under normal and faulty conditions: equations of magnetic circuit*

Let us analyze the operation of the device using equations of magnetic circuits for an ideal transformer. The transformer is ideal in the sense that its core is lossless, there are no leakage fluxes and no losses in the windings. Assume also that the reluctances of the yokes are negligible and all the legs are identical. Then the equations of magnetic circuits with magnetomotive forces are:

$$\Phi_1 R + \Phi_2 R = I_1 w_1 + I_2 w_2 - I_2 w_3 + I_s w_s ,$$
$$\Phi_1 R + \Phi_3 R = I_1 w_1 + I_2 w_2 , \qquad (11)$$
$$\Phi_1 = \Phi_2 + \Phi_3$$

where $R$ is the leg reluctance, $I_1$, $I_2$ and $I_s$ are the currents in windings $w_1$, $w_2$ and $w_s$ respectively; other notations are shown in Fig. 5.

In the normal regime $\Phi_2 = 0$, $\Phi_3 = \Phi_1$ and Eqs. (11) determine the relationships between the currents:

$$I_2 = \frac{2\Phi_1 R}{w_2} - I_1 \frac{w_1}{w_2}, \qquad I_s = -\Phi_1 R \left( \frac{w_2 - 2w_3}{w_2 w_s} \right) - I_1 \frac{w_1 w_3}{w_2 w_s} . \qquad (12)$$

In no-load conditions, $I_2 = 0$, the magnetic flux in the core is related to the magnetizing current $I_M$ according

$$I_M w_1 = 2\Phi_1 R . \qquad (13)$$

In most conventional power transformers $I_M$ is less than 1% of the nominal current and may be neglected in the comparison with $I_1$. Therefore the first terms in the right hand parts of (12) can be omitted. Using the second equation (12) one can relate the critical current of the SC winding to the instantaneous value of the secondary current $I_{2act}$ at which the device is activated (the activation current):

$$I_c = I_{2act} \frac{w_3}{w_s} . \qquad (14)$$

Under a fault the secondary voltage equals to zero:

$$w_2 \Phi_1 - w_3 \Phi_2 = 0 . \qquad (15)$$

Assuming that after the S-N transition in the limitation regime the resistance in the SC winding is large enough so that the current in the superconducting winding $I_s \approx 0$, we obtain

$$I_1 = \frac{2\Phi_1 R(w_2^2 - w_2 w_3 + w_3^2)}{w_1 w_3^2} = I_M\left(\frac{w_2^2}{w_3^2} - \frac{w_2}{w_3} + 1\right), \quad (16)$$

$$I_2 = \frac{\Phi_1 R(-2w_2 + w_3)}{w_3^2} = \frac{I_M w_1}{2w_3}\left(1 - 2\frac{w_2}{w_3}\right), \quad (17)$$

$$\Phi_2 = \Phi_1 \frac{w_2}{w_3}, \quad (18)$$

$$\Phi_3 = \Phi_1\left(1 - \frac{w_2}{w_3}\right). \quad (19)$$

It follows from these equations:

1. Deep limitation (much lower than the nominal current) of both the primary and secondary currents is achieved practically for all the ratios $w_2/w_3$ except the case when $w_3 \ll w_2$. This case, however, is unacceptable because $\Phi_2$ becomes much larger than $\Phi_1$ (see (18)) that requires a sufficient increase in the size of the legs and yokes.

2. The smaller $w_3$, the lower critical current of a superconductor is required (see (14)). On the other hand, if $w_3 < w_2$, the size of the magnetic core has to be increased. If $w_3 = w_2$ the magnetic fluxes $\Phi_2 = \Phi_1$ and all the legs have to be of the same size. If $w_3 > w_2$, $\Phi_2 < \Phi_1$ and the cross-section of the right leg can be less than that of other legs.

3. At $w_3 = 2w_2$ the secondary current $I_2$ equals zero (the complete compensation).

4. If $w_2 = 0$, at the short-circuit $\Phi_2 = 0$. Even without one section of the secondary we have the deep limitation: $I_1 = I_M$ and $I_2 = w_1 I_M / 2w_3$. It leads to another design in which the secondary and short-circuited SC windings are placed on different side legs and the primary is on the central leg. A drawback of this design is a large leakage inductance in the nominal conditions. Note that, at the application of transformers with higher level of the leakage inductance, no additional measures of fault current limitation are needed.

Similar results can be obtained for an "inverse" transformer in which the primary and secondary circuits (and the directions of energy fluxes) are interchanged. In this case the voltage $U_{nom}$ is applied to the terminals of windings $w_2$ and $w_3$. Since a transformer is a bilateral device, the inverse design operates in the same way and is described by the same set of equations (11). Under the normal conditions, the device operates as a usual transformer with the primary $w_2$ and secondary $w_1$. Under a fault, the condition (15) changes by $\Phi_1 = 0$ and therefore

$$U_{nom} = \omega(w_2\Phi_1 - w_3\Phi_2) = -\omega w_3\Phi_2 , \qquad (20)$$

i. e. $\Phi_2$ changes in inverse proportion with $w_3$.

The currents in the windings (now $I_1$ is the current in the output connected with the winding $w_1$, and $I_2$ is the input current) become

$$I_1 = -\frac{\Phi_2 R(-2w_2 + w_3)}{w_1 w_3}, \quad I_2 = -\frac{2\Phi_2 R}{w_3} . \qquad (21)$$

These currents are of the order of the magnetizing current.

Eqs. 11 allow to analyze the basic peculiarities of the CLT operation. More accurate calculation using the equations of the multi-winding transformer [12, 14, 16] with account of leakage and non-zero reluctance of the yokes gives similar results. Even at 30% leakage of the windings located on different legs we obtain only a small change in the primary and secondary fault currents, and this change can be compensated by the increase of the turn number in winding $w_3$.

*C. Magnetic system*

While the parameters of an FCL are chosen from the requirements to a current limiting device, the basic parameters of a CLT have to be chosen from its operation as a usual transformer. As opposite to an FCL, the magnetizing inductance of a CLT is much more (~100 times) than $z_b = U_{nom}/I_{nom}$ and it cannot be object for a variation. The variable parameters can be only $R_n/x_{sc}$ and $w_3/w_2$ ($x_{sc}$ is the inductive reactance of the SC winding). As for a usual single-phase transformer with the power $S$, the core diameter $D_{core}$ and the turn number $w_1$ in the primary winding are determined by the expressions [23]:

$$D_{core} = K_{tr} \sqrt[4]{\frac{S}{u_x}}; \quad w_1 = K_w \frac{U_{nom}}{D_{core}^2} , \qquad (22)$$

where $u_x$ is the short-circuit voltage of the transformer in percents of $U_{nom}$. The coefficients $K_{tr}$ and $K_w$ depend on the design features of the windings and magnetic core, isolation gaps, maximum magnetic flux density in the core; they are evaluated for a transformer with $S$ = 10-30 MVA as $K_{tr} \approx 4 \cdot 10^{-3}$ and $K_w \approx 3 \cdot 10^{-3}$ in units of SI [23]. The voltage $u_x$ in conventional transformers is determined by leakage flux between the primary and secondary windings and is of the order of 10%. It is possible to design a power transformer with lower $u_x$ but it results in the increase of short-circuit currents. In the CLT, $u_x$ contains two components related to the leakage between $w_1$ and $w_2$, and to the leakage between $w_3$ and $w_s$. The sum of these components has not to exceed the short-circuit voltage of a usual transformer. Due to current-limiting property of an CLT, the total value of $u_x$ can be essentially less. Note that in Eq. (22) one has to substitute the component of $u_x$ related to the windings located on the same leg.

The core diameters for the central and left legs and their yokes should be the same because in the normal conditions they carry the same flux. As shown above, the flux in the right leg and yoke is determined by the ratio $w_2/w_3$ (Eq. (18)); if $w_3 > w_2$ the diameter of the core in this part could be reduced.

*D. SC switching element*

The design of the switching element of a CLT is similar to that of the FCL.

Fig. 6 shows the secondary current and losses in the switching element as functions of the ratio $R_n/x_{sc}$. The curves were calculated using the equations of a multi-winding transformer [12, 14, 16] for two cases of the relationship between $w_3$ and $w_2$. The losses and temperature rise in the switching element were determined in the same approximation that for an FCL. As an example, it was chosen 15 MVA, 20 kV/6.3 kV single-phase transformer with the total value of the short-circuit voltage $u_x$ = 10%. The calculation using (22) gives $D_{core}$ =0.62 m, $w_1$ =155. In contrast to an inductive FCL, even the small ratio of $R_n/x_{sc}$ leads to a deep current limitation, especially in the case of the complete compensation $w_3 = 2w_2$.

Fig. 7 shows that heating of the switching element is considerable only at very small values of $R_n/x_{sc}$. Therefore the problem of the transition of the element into the normal state becomes else more critical than in an FCL.

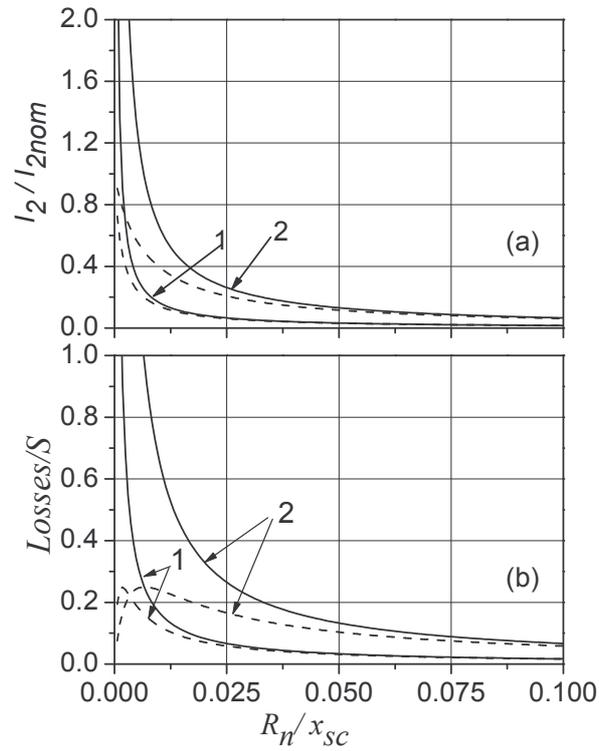

Fig. 6. Secondary current (a) and losses in the switching element (b) as functions of $R_n/x_{sc}$ during a fault in the secondary (solid lines) and after recovery of the normal regime in the circuit (dashed lines). Curves 1: $w_3 = 2w_2$, curves 2: $w_3 = w_2$.

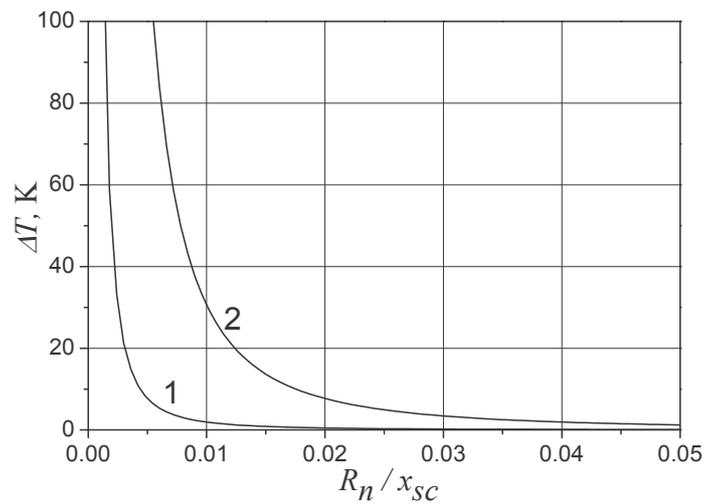

Fig. 7. Heating of a switching element after 0.1 s of the fault current limitation as a function of $R_n/x_{sc}$. Curve 1: $w_3 = 2w_2$, curve 2: $w_3 = w_2$.

## IV. Parameters of full-scale devices

*A. Application*

Let us evaluate the parameters of full-scale FCL and CLT for the same circuit (Fig. 8). First let the circuit contains a three-phase 35/11 kV, 45 MVA transformer. Inductive FCL can be installed in each phase (1) before the transformer at the high-voltage side, i. e. at point A (the phase parameters $U_{nom}$ = 20 kV, $I_{nom}$ = 750 A); (2) after the transformer at the low-voltage side, i. e. at point B ($U_{nom}$ = 6.3 kV, $I_{nom}$ = 2400 A); (3) in a line after the buses, at point C ($U_{nom}$ = 6.3 kV, $I_{nom}$ = 480 A). As an alternative, consider 3 single-phase 20/6.3 kV, 15 MVA CLTs, wye-wye connected replacing the three-phase transformer in Fig. 8.

We consider the application of HTS BSCCO hollow cylinders as SC short-circuited windings of an FCL and a CLT. The maximum resistance of the switching element is achieved if the whole of the cylinder passes into the normal state:

$$R_{n\max} = \frac{\rho_n \pi J_c D_{core}}{I_c} \qquad (23)$$

where the critical current of the element $I_c$ and $D_{core}$ are determined by (6) and (9) for an FCL and by (14) and (22) a CLT.

Thus, the resistance of the element is unequivocally determined by the core diameter.

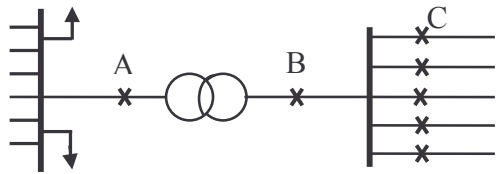

Fig. 8. Schema of the possible locations of FCLs in a distribution substation (the locations are marked by **x**).

Suppose the parameters of the cylinders: the normal state resistivity $\rho_n$ = $10^{-5}$ Ω·m, the critical current $J_c$ = $2\times10^7$ A/m² , the critical temperature $T_c$ = 105 K, and the specific volume heat capacity $C_V = 0.7 \times 10^6$ J/(m$^3$K). It is assumed that cooling is provided by liquid nitrogen at $T = 77$ K.

*B. FCL employing an HTS hollow cylinder*

Figs. 9 and 10 present the calculation results for two FCL designs installed in a network as shown in Fig. 8. The activation current was chosen equal to the minimum: $2\sqrt{2}I_{nom}$. For the calculation of the closed core design, it was taken $x_s = 5\% \, z_b$. This allows us to obtain a single-valued solution for the FCL parameters. The plots are built in the normalized units with the base units $U_{nom}$, $I_{nom}$, $z_b$ and $S$ referred to the circuit where the FCL is installed. In these units, the results for the FCLs in the transformer feeder from low voltage and high voltage sides coincide.

The inductive reactance of an FCL $x_n$ can be varied in a wide range adjusting to the desired level of the fault current limitation. However, the different core designs have the different range of the possible change of $x_n$. The application of the open core design is restricted by the condition $x_n / z_b < 1$: at higher values, the leakage reactance $x_s$ becomes unacceptable high. Therefore, this design cannot be used for the deep current limitation, $I_L < I_{nom}$. The minimal limited current, which can be achieved using the closed core design with the cylinder, is about $0.7 I_{nom}$ for an FCL in the transformer feeder and $0.4 I_{nom}$ for an FCL in line (Fig. 9a). Emphasize, that this design is unrealizable for small $x_n$ because the air gap in the magnetic core increases with the decrease of $x_n$ and becomes unrealizable large.

The considered designs have the different application ranges but, as Fig. 9a shows, the limited current depends on the design slightly. By contrast, the losses and temperature rise manifest a strong dependence on the design and installation place (Fig. 9b, c).

The rise of the cylinder temperature (Fig. 9c) is estimated in an adiabatic approximation assuming that a homogeneous cylinder fast passes into the normal state and keeps this state during the limitation time 0.1 s. The final temperature of the cylinder after fault clearing (recovery of the nominal load in circuit) was calculated as the steady-state temperature at $t \to \infty$ (Fig. 10).

For open core design, the requirement of the heating of the cylinder above the critical temperature are fulfilled only at $x_n/z_b < 0.45$ for the installation into a transformer circuit and at $x_n/z_b < 0.1$ for the installation into a line. At higher $x_n/z_b$, the cylinder resistance is lower than the normal state resistance $R_n$. This conclusion was proved experimentally in [18].

Fig. 10 shows that the recovery of the SC state in the cylinder without non-current pause is possible only at small $x_n/z_b$.

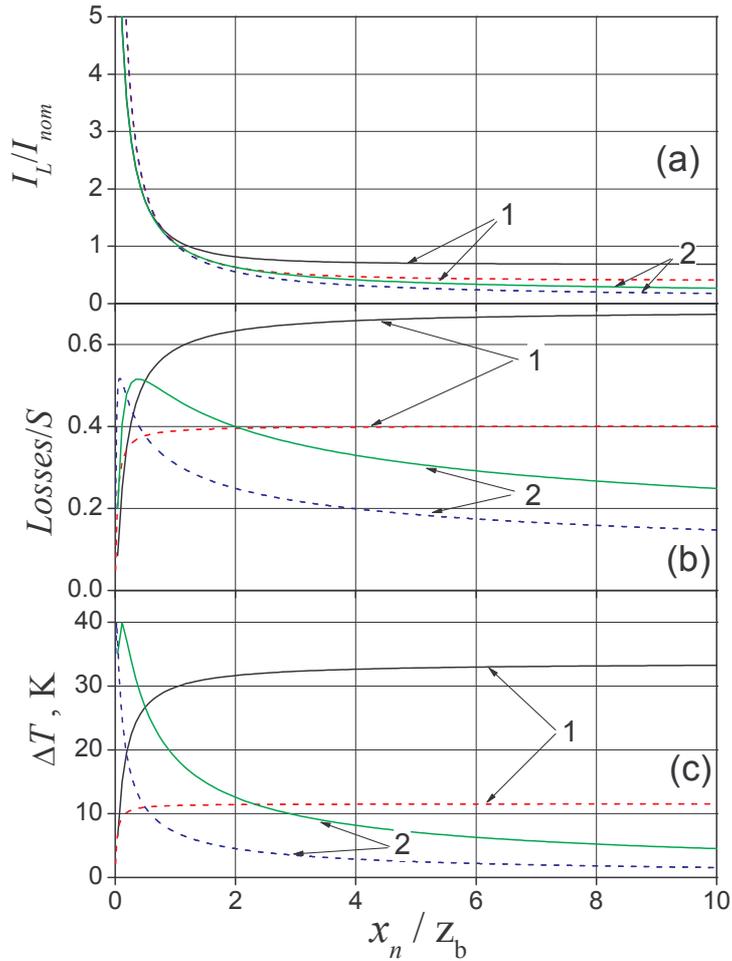

Fig. 9. Limited current (a), losses (b) and temperature increase (c) as functions of $x_n/z_b$ for an inductive FCL employing an HTS cylinder. Solid lines - FCL in transformer feeder; dashed lines - FCL in a line (see Fig. 8). 1- closed core design; 2 – open core design. $I_{nom}$, $z_b$ and $S$ are referred to the circuit where an FCL is installed.

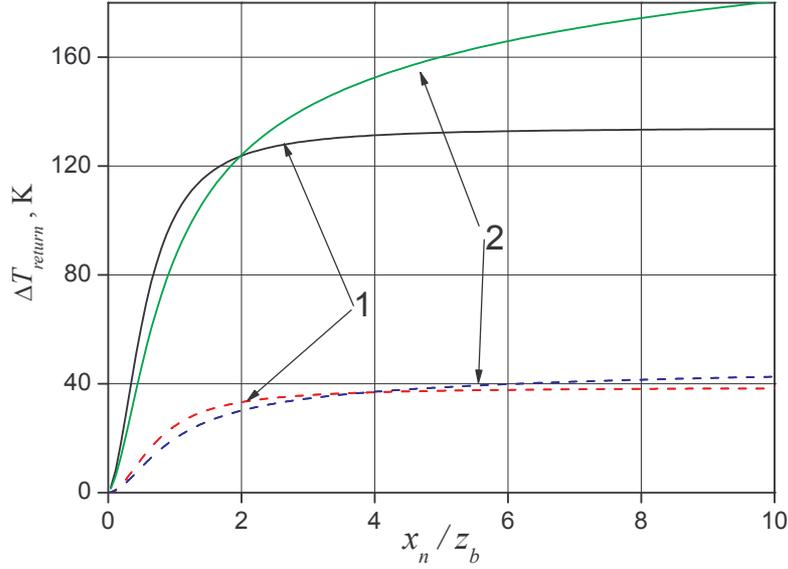

Fig. 10. Heating of the cylinder at the recovery of the nominal load in a circuit (t → ∞). Solid lines - FCL in transformer feeder; dashed lines - FCL in a line (see Fig. 8). 1- closed core design; 2 – open core design.

*C. CLT with an HTS cylinder*

The core diameter and turn numbers $w_1$ and $w_2$ are uniquely determined from (22). Simple calculations with the secondary voltage $U_{2nom}$ =6.3 kV, current $I_{2nom}$ = 2400 A and $w_3 = 2w_2 = 98$ give for the BSCCO cylinder $R_n = 5.9 \times 10^{-4}$ Ω and $R_n/x_{sc}$ = 0.0035 . Fig. 11 shows how the secondary current, losses and temperature change with the turn number $w_3$ within the limits $w_2 < w_3 < 2w_2$ . One can see that heating above the critical temperature is possible only when $w_3$ < 70 and for any $w_3$ the cylinder cannot be cooled enough in the conditions of the nominal load. The calculations show that the losses in the cylinder fall tenths times during a non-current pause in the secondary circuit.

*D. Performance example*

As an example, the basic parameters of the HTS cylinder based CLT and FCL installed in a transformer feeder (low voltage side) are summarized in Table 1 for several different values of the limited current.

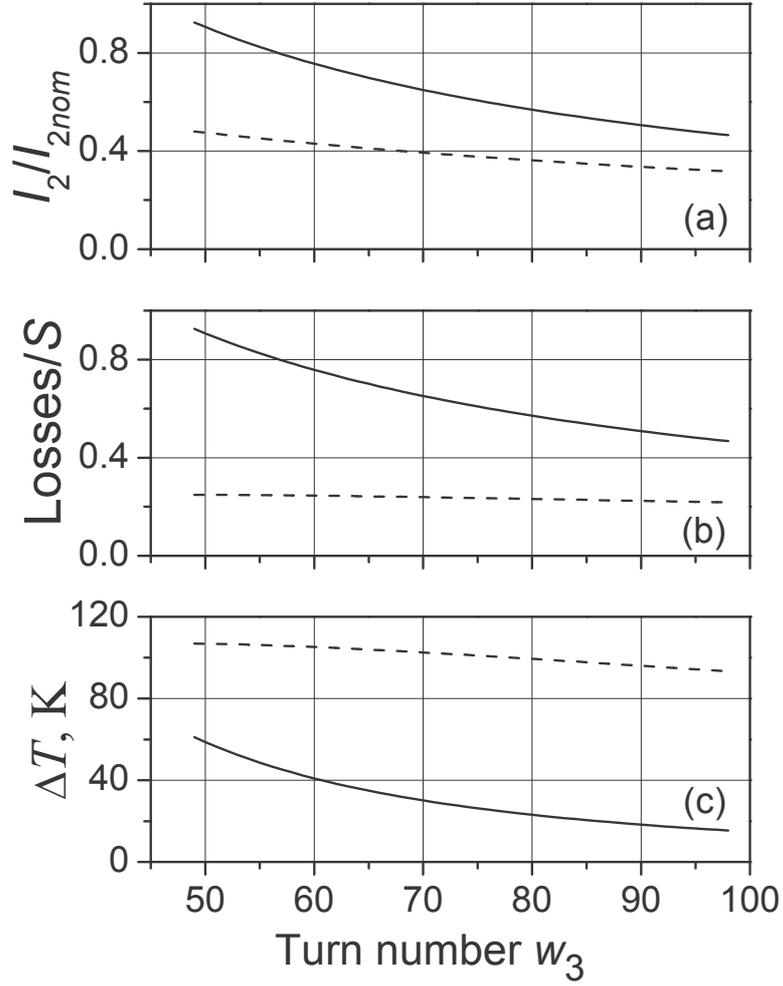

Fig. 11. Secondary current (a), losses (b) and increase of temperature (c) as functions of $w_3$. Solid lines – short-circuit in the secondary, dashed lines – recovery of the normal regime in the circuit.

**V. Discussion**

The proposed simple expressions allow to evaluate the basic parameters of an inductive FCL and a CLT and to determine their application areas. Also, the calculations show the feasibility of using HTS cylinders as SC switching elements of the devices. Both devices, FCL and CLT, limit successfully fault currents even in cases when the switching element resistance is sufficiently lower than the reactance of the SC winding. However, at the same resistance of the switching element, the current limited by a CLT is 5-10 times lower than the one limited by an FCL (compare Figs. 2 and 6). As follows from Fig. 2, the required limited current can be achieved by variation of both the resistance of the switching element and the reactance $x_n$. As distinct from this, the variable parameters in a CLT are the resistance of the switching element and the ratio of turn numbers in the two sections of the secondary winding $w_3/w_2$ while the magnetizing inductance remains

Table 1.

Parameters of FCL and CLT for different values of the limited current

| Parameters of electromagnetic system | | | | | |
|---|---|---|---|---|---|
| Device | FCL | | | CLT | |
| Design | Open core | Closed core | | Single-phase transformer | |
| Limited current | $2I_{nom}$ | $0.8 I_{nom}$ | $2I_{nom}$ | $0.92I_{nom}$ | $0.46I_{nom}$ |
| Inductive impedance, $\Omega$ | 1.1 | 16 | 1.1 | 4000 | 4000 |
| Primary turn number | 56 | 59 | 62 | 155 | 155 |
| Second. turn number $w_2$ | - | - | - | 49 | 49 |
| Turn number $w_3$ | - | - | - | 49 | 98 |
| Core diameter, m | 0.62 | 0.59 | 0.52 | 0.62 | 0.62 |
| Parameters of HTS hollow cylinder | | | | | |
| Volume, m$^3$ | 0.038 | 0.037 | 0.035 | 0.065 | 0.032 |
| Normal state resist., $\Omega$ | $1.1 \cdot 10^{-3}$ | $9.4 \cdot 10^{-4}$ | $8.0 \cdot 10^{-4}$ | $6 \cdot 10^{-4}$ | $1.2 \cdot 10^{-3}$ |
| Losses power, W | $7.7 \cdot 10^6$ | $1.1 \cdot 10^7$ | $9.8 \cdot 10^6$ | $7.0 \cdot 10^6$ | $1.4 \cdot 10^7$ |
| Temperature rise, K | 29 | 32 | 25 | 61 | 15 |
| Temperature at nominal load recovery, K | 37 | 124 | 50 | 105 | 95 |

unchanged. In principle, one can obtain a high limited current also with a CLT using low resistance switching elements but it leads to high losses and overheating of the element (Figs. 6 and 7). Two different designs of an FCL – open and closed core- complement each other in the application area: the first is applicable for $I_L > I_{nom}$, the second one is intended for the deep limitation.

The deep current limitation decreases the losses and heating of the switching element. As a result, the element is not heated up to the critical temperature (Fig. 3 and Table 1). As shown in [12, 18], a limiting effect is achieved also at the incomplete transition of the element into the normal state (resistive state) or at the transition of the part of the element, i. e. when the resistance is lower than the normal state resistance. In this case, the initial stage of the transition is characterized by higher limited current and losses in the superconductor. More exact calculation of the losses with taking into account of the transient process in the circuit and the resistive state of the superconductor gives higher values of the temperature increase [14].

Much deeper problem is the recovery of the SC state in the switching element after a fault clearing. There are two different regimes of the recovery of the SC state in the switching element: (1) under the nominal current in the circuit; (2) during a non-current pause. For example, if the fault occurs in the line with the FCL and a circuit-breaker opens the damaged circuit, the recovery happens under non-current conditions. In this case the power of the cryogenic equipment and the design of the winding should be chosen so to provide cooling down operating temperature before re-closure of the circuit. In the other case, when the FCL operates in an undamaged line, a fault clearing leads to the recovery of the nominal regime, and the FCL should return to the initial state at the nominal current in the circuit. In this case, for an FCL employing an HTS cylinder, the recovery is possible only at very small values of $x_n/z_b$ and, hence, large $I_L$ (Figs. 9 and 10). As one can see from Table 1, the CLT design based on a HSC cylinder provides the deep current limitation (below $I_{nom}$) for all real relationships between $w_2$ and $w_3$ . The recovery of the SC state in the cylinder of a CLT is possible only within a sufficient non-current pause.

Note, that, for both devices, the deep current limitation is problematic also from point of view of the stability of the SC state. As follows from Table 1, the normal state in the cylinder can exist at currents below the nominal value whereas the activation current is above the nominal one. Thus, at currents below the nominal currents, two stable states can exist at the same current in the protected circuit [24].

**VI. Conclusion**

We have shown that both SC devices, an FCL and a CLT can be successfully used for the solution of the fault current problem in power systems. A CLT allows deeper limitation of the current and can be more suitable for operation together with other superconducting devices, especially, with cables.

One obstacle for building conventional transformers with lower leakage reactance is the necessity to provide an admissible level of fault currents. Unique properties of the CLT- possibility to limit both the transient and steady-state currents - allow one to build transformers with low leakage reactance.

The problems of the stability in the SC state, the heating at the activation and recovery of the SC state after a fault clearing are most important issues for the operation of SC switching elements in both devices. It was shown that, especially for a CLT, even low values of the switching element resistance are sufficient to limit deeply fault currents. Therefore, one possible solution for the transition and recovery problems is to keep the switching elements in the resistive state (i. e. at the temperature below the critical one and at the current above the critical value). Another way is to fabricate the switching elements with high critical current density, for example, to use thin-film elements.

**References**

[1] R. F. Giese, and M. Runde "Assessment study of superconducting fault-current limiters operating at 77 K," *IEEE Trans. Power Delivery*, vol. 8(7), pp. 1138-47. 1993.

[2] E. Leung "Surge protection for power grids," *IEEE Spectrum*, vol. 34, pp. 26-30. 1997.

[3] V. Sokolovsky, V. Meerovich, I. Vajda, and V. Beilin "Superconducting FCL: design and application," *IEEE Trans. Appl. Supercond.*, vol. 14, pp. 1890-1900. Sept. 2004.

[4] Roadmap for Europe (2/06/01), in *Proc. Fault Current Limiter Working Group Workshop, SCENET*, Grenoble, CNRS, pp. 5-28, May 2001.

[5] Y. N. Vershinin, V. M. Meerovich, I. E. Naumkin, N. L. Novikov, and V. L. Sokolovsky, "A comparative analysis of nonlinear reactors with shields and high-

temperature (about 90 K) superconductors," *Elec. Technol. USSR* (*UK*), no. 1, pp. 1-9, Jan. 1989.

[6] V. Sokolovsky, V. Meerovich, V. Beilin, I. Vajda, "Application of an HTS thin film switching element in the inductive current limiter", *Physica C*, vol. 386, pp. 480-484, 2003.

[7] V. Meerovich, V. Sokolovsky, G. Jung, and S. Goren, "High-Tc superconducting inductive current limiter for 1 kV/25A performance," *IEEE Trans. Appl. Superconduct.*, vol. 5, pp. 1044-1048, June 1995.

[8] W. Paul, M. Lakner, J. Rhyner, P. Unternahrer, Th. Baumann, M. Chen, L. Windenhorn, and A. Guerig, "Test of a 1.2 MVA high-Tc superconducting fault current limiter," in Inst. Phys. Conf. Ser., vol. 158, pp. 1173-1178, 1997.

[9] C. Kurupakorn, N. Hagakawa, N. Kashima, S. Nagaya, M. Noe, K.-P. Juengst, and H. Okubo, "Development of high temperature superconducting fault current limiting transformer (HTc-SFCLT) with Bi2212 bulk coil,", *IEEE Trans. Applied Supercond.*, vol.14, no. 2, pp.900-903, June 2004.

[10] J. Noudem, J. M. Barbut, O. Sanchez, P. Tixador, and R. Tournier, "Current limitation at 1080 A under 1100V with bulk Bi-2223," *IEEE Trans. Appl. Supercond.*, vol. 9, pp. 664-67. June, 1999.

[11] Yu. A. Bashkirov, I. V. Yakimets, L. S. Fleishman, and V. G. Narovlyanskii, "Application of superconducting shields in current-limiting and special-purpose transformers," *IEEE Trans. Appl. Supercond.*, vol. 5, pp. 1075-1078, July 1995.

[12] V. Meerovich, V. Sokolovsky, "Experimental study of a transformer with superconducting elements for fault current limitation and energy redistribution," *Cryogenics*, vol. 45, pp.572-577, 2005.

[13] A. Gyore, G. Peter, and I. Vajda, "System investigation of high temperature superconducting self-limiting transformer," J. Physics: Conf. Series 43, pp. 966-970. 2006.

[14] V. Meerovich, V. Sokolovsky, and I. Vajda, " Comparison of a self-limiting transformer and a transformer type FCL with HTS elements," accepted to *IEEE Trans. Appl. Supercond.*, 2006 (presented at ASC'2006).

[15] I. Vajda, A. Gyore, S. Semperger, A. E. Baker, E. F. H. Chong, F. J. Mumford, V. Meerovich, and V. Sokolovsky, " Investigation of high temperature superconducting self-limiting transformer with YBCO cylinder," accepted to *IEEE Trans. Appl. Supercond.*, 2006 (presented at ASC'2006).

[16] L. V. Leites, "*Electromagnetic calculations of transformers and reactors*," Energiya, Moscow, 1981.

[17] V. Sokolovsky, V. Meerovich, G. Grader, and G. Shter, "Experimental investigation of current-limiting device model based on high-Tc superconductors," *Physica C*, vol. 209, pp. 277-280, 1993.

[18] V. Meerovich, V. Sokolovsky, J. Bock, S. Gauss, S. Goren, and G. Jung, "Performance of an inductive fault current limiter employing BSCCO superconducting cylinders," *IEEE Trans. Appl. Superconduct.*, vol. 9, pp. 4666-4676, December 1999.

[19] B. N. Neklepaev, "*Coordination and optimization of fault current level in power systems*," Moscow, Energya, 1978 (in Russian).

[20] V. M. Meerovich and V. L. Sokolovsky, "Calculation of electrical reactors without yokes," *Electrotechnica*, no. 10, pp. 31-33, 1983.

[21] C. W. T. McLyman, "*Transformer and Inductor Design Handbook* ", Marcel Dekker, Inc., 1988. 432p.

[22] O. Kiltie, "*Design Shortcuts and Procedures for Electronic Power Transformers and Inductors*", 2nd edition, Ordean Kiltie, 1981, 274p.

[23] P. M. Tichomirov, "*Calculation of transformers*," Energoatomizdat, Moscow, 1986, 528 p.

[24] V. M. Meerovich and V. L. Sokolovsky, S. Goren, and I. Vajda," *IEEE Trans. Appl. Superconduct.*, vol. 11, pp. 2110-2113, March 1999.